\documentclass[epjST]{svjour}

\usepackage{latexsym,exscale}
\usepackage[ansinew]{inputenc}
\usepackage{amsmath}
\usepackage{url}
\usepackage{textcomp}
\usepackage{amsfonts}
\usepackage{amssymb}
\usepackage{mathrsfs}
\usepackage{psfrag}
\usepackage{color} 
\usepackage{graphicx}
\usepackage{floatrow} 
\usepackage{pstricks}
\usepackage{subfigure}
\usepackage{anysize}
\usepackage[T1]{fontenc}
\usepackage{float}
\usepackage{booktabs}
\usepackage{geometry}
\usepackage{fancyhdr}
\usepackage{bm}

\usepackage{lmodern}						
\usepackage[english]{babel}

\begin{document}

\title{Excitable elements controlled by noise and network structure}

\author{B. Sonnenschein\inst{1,2}\fnmsep\thanks{\email{sonne@physik.hu-berlin.de}}
\and M. A. Zaks\inst{3}
\and A. B. Neiman\inst{4}
\and L. Schimansky-Geier\inst{1,2}}

\institute{Department of Physics, Humboldt-Universit\"at zu Berlin, Newtonstrasse 15, 12489 Berlin, Germany 
\and Bernstein Center for Computational Neuroscience Berlin, Philippstrasse 13, 10115 Berlin, Germany 
\and Institute of Mathematics, Humboldt-Universit\"at zu Berlin, Rudower Chaussee 25, 12489 Berlin, \\Germany
\and Department of Physics and Astronomy, Ohio University, Athens, Ohio 45701, USA}


\abstract{
We study collective dynamics of complex networks of stochastic excitable elements, active rotators. In the thermodynamic limit of infinite number of elements, 
we apply a mean-field theory for the network and then use a Gaussian approximation to obtain a closed set of deterministic differential equations. 
These equations govern the order parameters of the network. We find that a uniform decrease in the number of connections per element in a homogeneous network
merely shifts the bifurcation thresholds without producing qualitative changes in the network dynamics. In contrast, 
heterogeneity in the number of connections leads to bifurcations in the excitable regime. 
In particular we show that a critical value of noise intensity for the saddle-node bifurcation decreases with growing connectivity variance. 
The corresponding critical values for the onset of global oscillations (Hopf bifurcation) show a non-monotone dependency on the
structural heterogeneity, displaying a minimum at moderate connectivity variances.
}

\maketitle

\section{Introduction}
\label{intro}
Excitable elements are ubiquitous in physical, geophysical, chemical and biological systems. 
Examples range from chemical reactions, neuronal systems, cardiovascular tissues and climate dynamics 
to laser devices, see Ref. \cite{LiGarNeiSchi04,AniAstNeVaSchi07} and references therein. An excitable system possesses a stable equilibrium,
but, once excited by sufficiently strong perturbations, displays large non-monotonic
excursions before returning to the state of rest. 

Networks of excitable elements are omnipresent in nature, and  
the human brain, as a gigantic network of neurons, appears to be paradigmatic 
in this respect (cf. \cite{Sp10}). In networks of nonlinear oscillators, one typically encounters the collective phenomenon of 
synchronization: different oscillators adjust their rhythms due to 
coupling \cite{AniVadPoSaf92,PikRosKu03,BalJaPoSos10}. Recently much effort has been invested into 
understanding the mechanisms which stand behind synchronization in complex networks of nonlinear oscillators and excitable systems, 
see Ref. \cite{ZaSaLSGNei05,ArDiazKuMoZh08} and references therein.

A crucial step in characterizing limit cycle oscillators was done by Kuramoto~\cite{Kur84}:
for a broad class of systems, he reduced description to dynamics of a single scalar variable,
the uniformly rotating phase. Since the origins of this approach can be traced back to the
biologically motivated Winfree model~\cite{Win67}, it appears to be a good candidate
for providing fundamental insights into the functioning of coupled biological oscillators.
Specifically, in the context of computational neuroscience various questions can be addressed by studying the
Kuramoto approach with suitable modifications \cite{cumin2007,BreaHeiDaff10}.  
In general, coupled limit-cycle oscillators can be approximated by the Kuramoto model 
as long as the interaction between them stays weak, 
so that each oscillator experiences only a small phase perturbation.
Relative simplicity of the Kuramoto model turns it into a convenient tool
for analytical explorations. In the frame of this model, characteristics of different transitions 
between desynchronized and synchronous states in large ensembles of oscillators can be 
obtained in closed form. This explains continuing interest in the Kuramoto model
among physicists, neuroscientists and mathematicians \cite{Str00,AcBoViRiSp05}.

Phase oscillators in their canonical form are not directly applicable to studies of
coupled excitable elements. In particular a Kuramoto oscillator possesses translational
invariance with respect to phase shifts: it rotates uniformly, and all phase values
are dynamically equivalent,  whereas for an excitable element  a certain event (e.g. spike) is singled out.
An ``active rotator'' -- a modification of the Kuramoto model which takes into account 
these properties of excitable elements -- was introduced by Shinomoto and 
Kuramoto in~\cite{ShiKur86}. 
Since then, the analysis was refined and certain additional aspects were 
investigated in \cite{KoLuLSG02,TesScToCo07,ChStr08,NaKo10,LafCoTo10}.
Most of the research considered the case of uniform global coupling in the ensemble;
only in a few works, active rotators were studied on a 
network with complex structure, see e.g. \cite{TesZanTo08}.
Here we study dynamics of networks of randomly connected stochastic active rotators and put emphasis on how
temporal fluctuations and the network structure influence the dynamics of the network as a whole. 

The paper is organized as follows. In section~\ref{model} we introduce the basic
model and briefly discuss its properties. In section~\ref{sec_MF} the mean-field 
approximation is developed: the Fokker-Planck equation for the time-dependent
probability distribution of the phases is derived and transformed by means of
the Fourier expansion into the infinite hierarchy of ordinary differential equations. 
In section~\ref{gaussian_approx} this hierarchy is brought to a closed form
with the aid of a Gaussian approximation. Our approach is exemplified for
regular and binary random networks in section~\ref{example}. 
Numerical simulation of the binary network 
corroborates the conclusions drawn from the analysis of the mean-field model. 

\section{Model}
\label{model}

In the population of noise-driven active rotators,   
introduced by Shinomoto and Kuramoto \cite{ShiKur86},
dynamics of individual phases $\phi_i(t)$ is governed by equations
\begin{equation}
  \dot{\phi}_i(t)=1-a\sin\left(\phi_i(t)\right)+\frac{\kappa}{N}\sum_{j=1}^{N}A_{ij}\sin\left(\phi_j(t)-\phi_i(t)\right)+\xi_i(t),
\label{ourmodel}
\end{equation}
where the units are indexed by $i=1,\ldots,N$. 
Parameter $a$ determines the excitation threshold of each separate element, 
whereas $\kappa$ characterizes the coupling strength. 
Here, we restrict ourselves to the case where these parameters
are identical for all elements of the ensemble. 
The network is assumed to be undirected and unweighted, hence
the elements of the symmetric adjacency matrix $A$ adopt only two values:
$A_{ij}=1$, if the units $i$ and $j$ are coupled, and $A_{ij}=0$ otherwise. 
The number of connections of a node, given by
\begin{equation}
k_i=\sum_{j=1}^{N} A_{ij},
\label{degrees}
\end{equation}
is called the degree of this node. In absence of isolated nodes, $k_i>0$. 
Complex topologies of real-world networks can be encoded into the adjacency
matrix, and decoded by counting all the degrees, which gives rise to a degree distribution $P(k)$.
Throughout this paper we discuss random networks.

The terms $\xi_i(t)$ in (\ref{ourmodel}) are statistically independent across the network nodes and are zero mean Gaussian 
white noise sources,
$
  \langle\xi_i(t)\rangle=0,\;\;\;\;
\langle\xi_i(t)\xi_j(t')\rangle=2D\delta_{ij}\delta(t-t'),
$
where the angular brackets denote averages over different realizations of the noise and $D$ is the noise intensity.

For an isolated element in the absence of noise, $\dot{\phi}_i=1-a\sin\left(\phi_i\right)$, 
excitable behavior is apparent. For $\left|a\right|>1$ 
the stable equilibrium is  at $\phi_i^{\infty}=\arcsin(a^{-1})$,
and the system needs a sufficiently strong perturbation in order to make an excursion around
the circle. Noise  plays this role driving the system 
to escape from the equilibrium state $\phi_i^{\infty}$. An escape event corresponds  
to the release of a single spike \cite{ShiKur86,KurSchu95,LiGarNeiSchi04}.
For $\left|a\right|<1$ the element shows oscillatory
behavior with frequency $\sqrt{1-a^2}$. Notably, evolution of the phase $\phi$ is not uniform in this
case: it is at the slowest near $\phi=\pi/2$ and at the fastest near $\phi=3\pi/2$.

\section{Mean-field theory}
\label{sec_MF}
\begin{figure}
\centering
\includegraphics[width=0.5\linewidth]{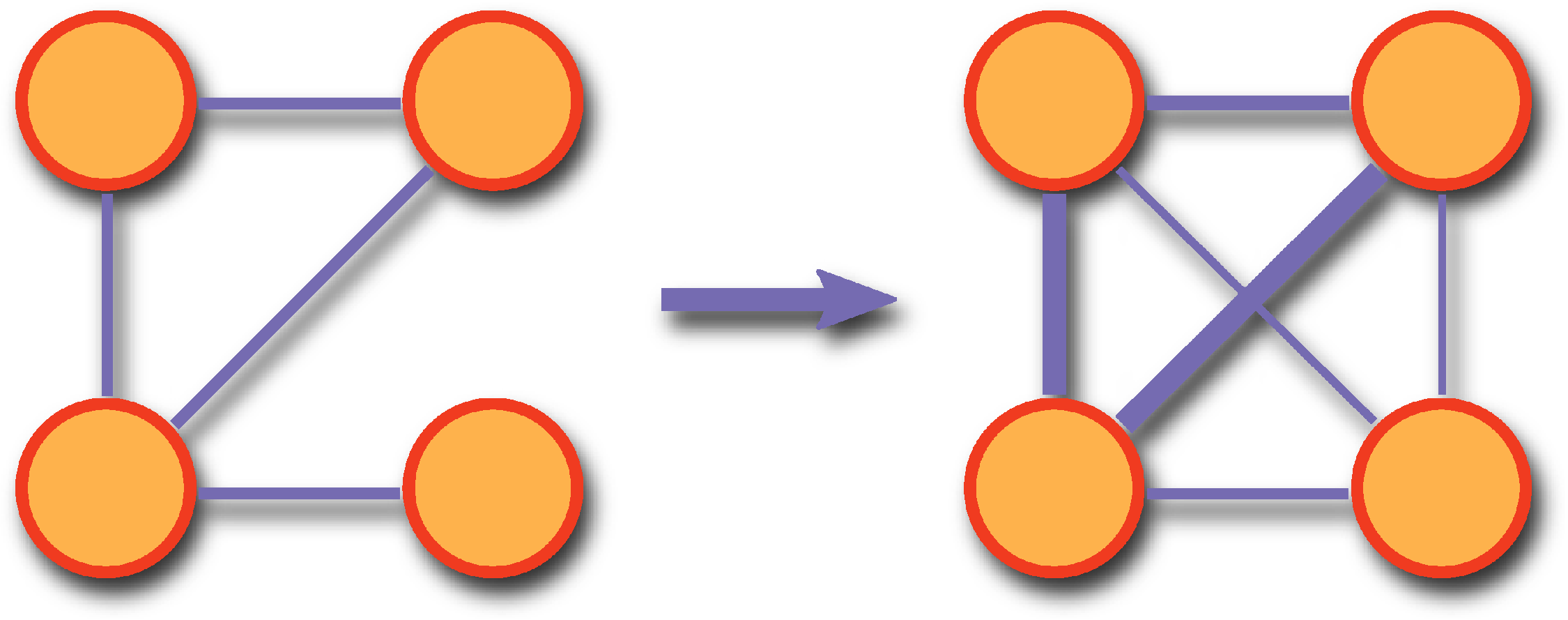}
\caption[]{(Color online) Effect of the mean-field
approximation on an undirected network of four nodes. Left: original complex network.
 Right: the approximate network. Thickness of the edges is approximately proportional to their weight.}
\label{MF}
\end{figure}

A mean-field description can be constructed by assuming that all nodes in the network 
are statistically identical, so that they are  fully characterized by their individual degrees \cite{SonnSchi12}. 
The derivation of the mean-field description can be done by coarse-graining the network structure, 
namely by replacing the original network by a fully connected one with random coupling
weights that mimic the actual connectivity (see Fig. \ref{MF} for illustration). 
Considering an uncorrelated random network \cite{Ichi04,SonnSchi12,SonnSagSchi13} and
requiring the conservation of the individual degrees, $k_i=\sum_{j=1}^{N} \tilde A_{ij},\ i=1,\ldots,N$, 
the elements of the approximate adjacency matrix read
\begin{equation}
  \tilde A_{ij}\,=\, k_i \, \frac{k_j}{\sum_{l=1}^{N} k_{l}}.
\label{newA}
\end{equation}
Inserting this into Eq. \eqref{ourmodel} yields a weighted mean-field approximation and effectively 
provides a one-oscillator description.

The system of $N$ coupled Langevin equations  \eqref{ourmodel} can be characterized in terms of the transition probability density
$\mathcal P_N\left(\boldsymbol{\phi},t|\boldsymbol{\phi}^0,t^0;\boldsymbol{k}\right)$. 
Here, the vector $\boldsymbol{\phi}=\left(\phi_1,\ldots,  \phi_N\right)$ is built
from the phases of $N$ rotators at time $t$ whereas
$\boldsymbol{\phi}^0=\left(\phi_1^0,\ldots, \phi_N^0\right)$ consists of their
values at initial time $t^0$. 
Components of the time-independent vector $\boldsymbol{k}=\left(k_1,\ldots,
  k_N\right)$ are the degrees of all nodes in the network.
We assume that the initial phases $\phi_i^0$ are identically and independently distributed.
The transition probability density obeys a Fokker-Planck equation (FPE) which
describes the evolution of the ensemble from time $t^0$ to time $t>t^0$.

Integration of the FPE over the $N-1$ phases $\phi_2,\ldots,\phi_N$, degrees $k_2,\ldots,k_N$ and corresponding initial phases \cite{SonnSchi12},
yields an evolution equation for the one-oscillator probability density $\mathcal P_1(\phi_1,t|k_1)$, 
in which, however, the two-oscillator density $\mathcal P_2(\phi_1,\phi_2,t|k_1,k_2)$ is present as well
(see Ref. \cite{CraDa99} for the case of all-to-all coupling with more general coupling functions):
\begin{equation}
\begin{aligned}
  \frac{\partial\mathcal P_1}{\partial t}=&-\frac{\partial}{\partial \phi_1}(1-a\sin\phi_1)\mathcal P_1 + D\frac{\partial^2\mathcal P_1}{\partial\phi_1^2} \\
  &- \frac{\partial}{\partial\phi_1}\left[\frac{\kappa}{N}\frac{k_1\left(N-1\right)}{\sum_{l}k_l}\int\mathrm d \phi_2\sum_{k_2}\sin\left(\phi_2-\phi_1\right)k_2\mathcal P_2\left(\phi_1,\phi_2,t;k_1,k_2\right)\right]\ .
\end{aligned}
  \label{two-osc}
\end{equation}
In the thermodynamic limit $N\rightarrow\infty$ the ratio $\sum_{l}k_l/(N-1)$
approaches the average degree $\langle k'\rangle$. In the framework of the mean-field approximation, 
correlation of phases between any two oscillators can be discarded,
$
  \mathcal P_2\left(\phi_1,\phi_2,t;k_1,k_2\right)\Rightarrow\mathcal P_1\left(\phi_1,t;k_1\right)\mathcal P_1\left(\phi_2,t;k_2\right)\,.
$
In this way Eq. \eqref{two-osc} becomes closed in $\mathcal P_1(\phi_1,t|k_1)$, 
yielding an effective one-oscillator description; the price is, that it becomes
a {\em nonlinear} FPE \cite{Fr05}:
\begin{equation}
  \frac{\partial\mathcal P_1\left(\phi,t|k\right)}{\partial
    t}\,=\,-\,\frac{\partial}{\partial\phi}\left[v(\phi,t|k) \, \mathcal P_1\left(\phi,t|k\right)\right]\,+\,D\frac{\partial^2
    \mathcal P_1\left(\phi,t|k\right)}{\partial\phi^2}.
\label{fpe}
\end{equation}
Nonlinearity enters this equation through the mean increment of the phase per unit time, 
$v(\phi,t|k)\equiv1-a\sin\left(\phi(t)\right)+r(t)\kappa k N^{-1}\sin\left(\Theta(t)-\phi(t)\right)$, which
depends on the density $\mathcal P_1(\phi,t|k)$ via the mean-field amplitude $r(t)$ and phase $\Theta(t)$, 
\begin{equation}
  r(t)\mathrm e^{i\Theta(t)}=\frac{1}{\langle
    k'\rangle}\int_{0}^{2\pi}\mathrm d\phi'\sum_{k'}\ \mathrm
  e^{i\phi'}\ \mathcal P_1\left(\phi',t|k'\right)\ k'\ 
  P\left(k'\right)\,,
  \label{order_fpe}
\end{equation}
where $P(k')$ denotes the degree distribution. Since all oscillators are statistically identical in this setting, from now on we 
neglect the indices for the individual phases and degrees.
We note that Eq. \eqref{order_fpe} suggests to introduce order parameters for each degree class, namely
\begin{equation}
 r(t)\mathrm e^{i\Theta(t)}\equiv\frac{1}{\langle k'\rangle}\sum_{k'}\ r_{k'}(t)\mathrm e^{i\Theta_{k'}(t)}\ k'\ P\left(k'\right)\, .
 \label{rkthetak}
\end{equation}

Let us proceed with the Fourier series expansion of the one-oscillator probability density:
\begin{equation}
\mathcal P_1(\phi,t|k)=\frac{1}{2\pi}\sum_{n=-\infty}^{+\infty}\rho_n(t|k)\mathrm e^{-in\phi}\ ,
\label{expansion} 
\end{equation}
with $\rho_0=1$ and $\rho_{-n}=\rho_{n}^*$.
Through the inverse transform of \eqref{expansion} one can write
\begin{equation}
\rho_n(t|k)\equiv\frac{1}{N_k}\sum_{j\in\{k\}}\mathrm e^{i n\phi_j}\equiv c_n(t|k)+is_n(t|k),
\label{orderp}
\end{equation}
where $n=1,2,\ldots,\infty$; 
$N_k$ stands for the number and $\{k\}$ for the set of nodes with degree $k$. 
For every degree $k$ that appears in the network, the Fourier coefficients $\rho_n(t|k)$
are governed by an infinite chain of coupled complex-valued equations
\begin{equation}
\begin{aligned}
\frac{\dot\rho_n(t|k)}{n}=&\frac{\tilde\kappa k}{\langle k'\rangle}\left(\rho_{n-1}(t|k)\sum_{k'}\rho_1(t|k')k'P(k')-\rho_{n+1}(t|k)\sum_{k'}\rho_{-1}(t|k')k'P(k')\right)\\
&+\frac{a}{2}\left(\rho_{n-1}(t|k)-\rho_{n+1}(t|k)\right)-(Dn-i)\rho_{n}(t|k),
\end{aligned}
\label{chain}
\end{equation}
with the abbreviation $\tilde\kappa:=\kappa/(2N)$.  The mean degree $\langle k'\rangle$ of
the network is the first moment of the degree distribution.
As expected, in the case of all-to-all connectivity the results of \cite{SakShiKur88}
are recovered. 
Equation \eqref{chain} provides a complete description of our system in the mean-field approximation, 
which is valid as long as $k\gg 1$. Thus, for a sparsely connected network the 
mean-field description is likely to fail \cite{ResOttHu05,SonnSchi12,PerErBaTirJe13}.
We also note that the above mean field approach is restricted to networks with finite 
second moment of the degree distribution \cite{ResOttHu05,Lee05,SonnSchi12}.

\section{Gaussian approximation}
\label{gaussian_approx}
There seems to be no gain in reduction of complexity from 
recasting the Langevin dynamics \eqref{ourmodel} into the mean field description \eqref{chain}. 
One approach to turn \eqref{chain} into a dynamical system of finite order
is to truncate the Fourier series. In the case of all-to-all coupling the values of  $|\rho_n|$
decay with increase of $n$ sufficiently fast, and the quantitative dynamics is correctly
captured by the first two dozens of coefficients so that the rest of the series can be neglected \cite{SakShiKur88,ZaNeFeSch03}. 
An alternative approach lies in finding a framework in which a closure of the infinite set of equations \eqref{chain} 
can be achieved. The so-called Ott-Antonsen ansatz \cite{OttAnt08} yields impressive results for deterministic ensembles
of coupled Kuramoto oscillators, reducing exactly the infinite-dimensional set of equations to a low-dimensional system.
Unfortunately, for the case of noisy systems the direct application of the Ott-Antonsen ansatz 
is impossible, and we are unaware of its appropriate modifications.
Instead, we use a Gaussian approximation (GA): we assume that in every subset of 
oscillators with the same network degree, the distribution of the phases at every moment of time
is Gaussian with the mean $m_k(t)$ and the variance $\sigma_k^2(t)$. 
The smaller the noise intensity $D$, the more accurate is the GA \cite{KurSchu95}. The 
noise should not dominate the collective dynamics, i.e. $D\lesssim\kappa\langle k\rangle N^{-1}$.
For a recent utilization of the GA, see Ref. \cite{SonnSchi13}.

In the thermodynamic limit, $N\to\infty$, the averaging in Eq. \eqref{orderp} in the GA  
yields 
\begin{equation}
c_n(t|k)=\mathrm e^{-n^2\sigma_k^2(t)/2}\cos\left(nm_k(t)\right),\ \ s_n(t|k)=\mathrm e^{-n^2\sigma_k^2(t)/2}\sin\left(nm_k(t)\right),
\label{cs}
\end{equation}
therefore all $c_n$ and $s_n$ are predetermined by $c_1$ and $s_1$: $c_2=c_1^4-s_1^4$, $s_2=2s_1c_1\left(s_1^2+c_1^2\right)$, etc. \cite{ZaNeFeSch03}.
Thus, the dimension of the reduced system of equations for the Fourier coefficients is  twice the number 
of different degrees present in the network. Denoting $c_1(t|k)$ by $c_1$ and $s_1(t|k)$ by $s _1$, Eqs. \eqref{chain} turn into
\begin{equation}
\begin{aligned}
\dot c_1&=\frac{\tilde\kappa k}{\langle k'\rangle}\left[\langle c_1(t|k') k'\rangle(1-c_1^4+s_1^4)-2 \langle s_1(t|k') k'\rangle s_1 c_1 (s_1^2+c_1^2)\right]+\frac{a}{2}(1-c_1^4+s_1^4)-D c_1-s_1,\\
\dot s_1&=\frac{\tilde\kappa k}{\langle k'\rangle}\left[\langle s_1(t|k') k'\rangle(1+c_1^4-s_1^4)-2 \langle c_1(t|k') k'\rangle s_1 c_1 (s_1^2+c_1^2)\right]-a s_1 c_1 (s_1^2+c_1^2)-D s_1+c_1,
\end{aligned}
\label{dotcs}
\end{equation}
for every degree $k$ in the network. Averaging $\langle\ldots\rangle$ is made over the degree distribution $P(k')$. 
Equivalently, equations for the mean $m_k(t)$ and the variance $\sigma_k^2(t)$ are
\begin{equation}
\begin{aligned}
\dot m_k=&1-\mathrm e^{-\sigma_k^2/2}\cosh\sigma_k^2\\
&\times\left[a\sin m_k-\frac{\kappa k}{N\langle k'\rangle}\left(\langle k'\mathrm e^{-\sigma_{k'}^2/2}\sin m_{k'}\rangle\cos m_k-\langle {k'}\mathrm e^{-\sigma_{k'}^2/2}\cos m_{k'}\rangle\sin m_k\right)\right],\\
\dot \sigma_k^2=&2D-2\mathrm e^{-\sigma_k^2/2}\sinh\sigma_k^2\\
&\times\left[a\cos m_k+\frac{\kappa k}{N\langle k'\rangle}\left(\langle k'\mathrm e^{-\sigma_{k'}^2/2}\sin m_{k'}\rangle\sin m_k+\langle {k'}\mathrm e^{-\sigma_{k'}^2/2}\cos m_{k'}\rangle\cos m_k\right)\right].
\end{aligned}
\label{dotmvar}
\end{equation}

\section{Results: regular and binary random networks}
\label{example}

The integro-differential character of Eqs. \eqref{dotmvar} makes further general analysis cumbersome. Below we discuss explicitly 
two relatively simple examples of  regular and binary random networks.

\subsection{Regular networks}

For regular networks where all nodes have the same degree Eqs. \eqref{dotmvar} simplify to two coupled equations:
\begin{figure}
\centering
\includegraphics[width=0.7\linewidth]{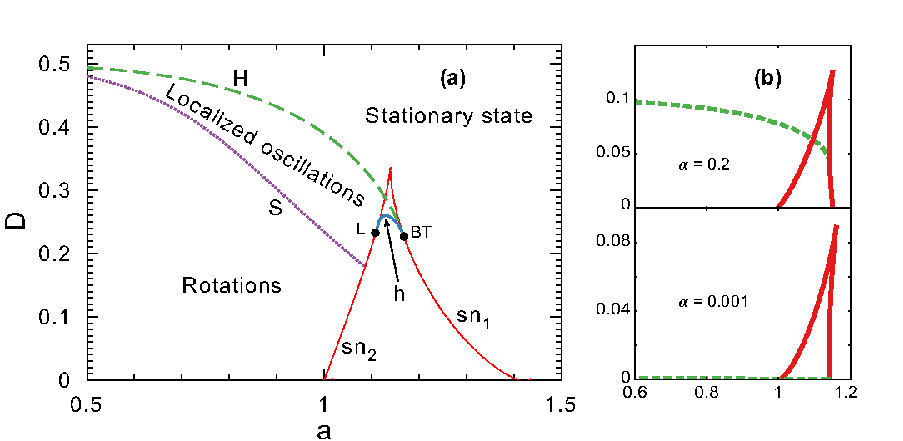}
\caption[]{(Color online) Bifurcation diagram of a regular network of stochastic rotators in the GA. 
(a): The coupling strength equals $\kappa=1$ and the connectivity fraction $\alpha$ is unity. $sn_{1,2}$: saddle-node bifurcation lines; $H$: Hopf bifurcation line; $h$: homoclinic bifurcation, $S$: oscillation type changes; $BT$: Bogdanov-Takens bifurcation point, $L$: homoclinic loop attached to the saddle-node. 
(b): Fragments of the bifurcation diagram for the indicated values of the connectivity fraction $\alpha$.}
\label{bif}
\end{figure}
\begin{equation}
\dot m_k=1-a\mathrm e^{-\sigma_k^2/2}\cosh\sigma_k^2\sin m_k,\quad
\dot \sigma_k^2=2D-2\mathrm e^{-\sigma_k^2/2}\sinh\sigma_k^2\left[a\cos m_k+\kappa kN^{-1}\mathrm e^{-\sigma_k^2/2}\right].
\label{dotmvar_reg}
\end{equation}
The only difference to the known cumulant equations for a fully connected network $k=N$ \cite{ZaNeFeSch03}
is a rescaling of the coupling strength: $\kappa\rightarrow\kappa k/N$ (compare panels (a) and (b) in Fig. \ref{bif}).
Figure \ref{bif} panel (a) shows a complete bifurcation diagram for all-to-all coupling on the parameter plane excitability $a$ and noise intensity $D$. 
Two saddle-node bifurcations, denoted by $sn_1$ and $sn_2$, separate parameter regions
in which only a single steady state exists from the central wedge-shaped region in which
three steady states are present. While at $sn_2$ a saddle point and a node are created, 
$sn_1$ destroys the saddle and the original steady state. 
On both bifurcation curves one of the Jacobian eigenvalues vanishes. There is a 
point where the second eigenvalue vanishes as well. This codimension-2 Bogdanov-Takens (BT) bifurcation 
is an origin of two further bifurcation lines: the 
Hopf bifurcation $H$ and the homoclinic bifurcation $h$. While above $H$, the steady state is stable, below $H$ the stable limit cycle comes into existence. The curve $h$
marks the existence of an orbit homoclinic to the saddle point. In the parameter region between the curves $H$, $h$ and $sn_2$ the stable steady state coexists
with the attracting limit cycle. In the region between $H$ and $sn_{1,2}$ two stable steady states exist. At the point $L$, saddle-node and homoclinicity merge together \cite{ZaNeFeSch03}.

Above the Hopf bifurcation line $H$, individual units are firing incoherently, and therefore do not create a globally synchronized oscillation. So in that stationary state, even though
$m_k$ and $\sigma_k$ are constant in time, the current $\sum_j\langle\dot\phi_j(t)\rangle_t/N$ does not vanish ($\langle\ldots\rangle_t$ stands for time average); it rather grows with 
further increase of $D$ until saturation \cite{TesScToCo07}. Below $H$, small-amplitude oscillations are born, where $m_k$ is localized between two phase values (``localized oscillations''). 
With decreasing noise intensity $D$ their amplitude grows. Below the line $S$, oscillations turn into full rotations: instead of oscillating back and forth, the ensemble
rotates around the whole circle $[0,2\pi)$. Upon the line $S$, $\sigma_k$ becomes infinite; below $S$ it is finite again. Finally, below the saddle-node bifurcation line, 
the system reaches again a stationary state, but this time the oscillators are mostly at rest without producing any current $\sum_j\langle\dot\phi_j(t)\rangle_t/N \sim 0$.

In the following, we do not discriminate between localized oscillations and full-circle rotations. We focus on how oscillating patterns are affected by the network structure. Therefore,
we concentrate on finding the Hopf and the saddle-node bifurcations.

Since a reduction in the number of connections effectively decreases the coupling strength, no new qualitative phenomena are expected if all nodes share the same degree. 
In what follows, we use 
the abbreviation $\alpha:=k/N$ for the connectivity fraction.

\begin{figure}
\centering
\begin{tabular}{cc}
\includegraphics[width=0.49\linewidth]{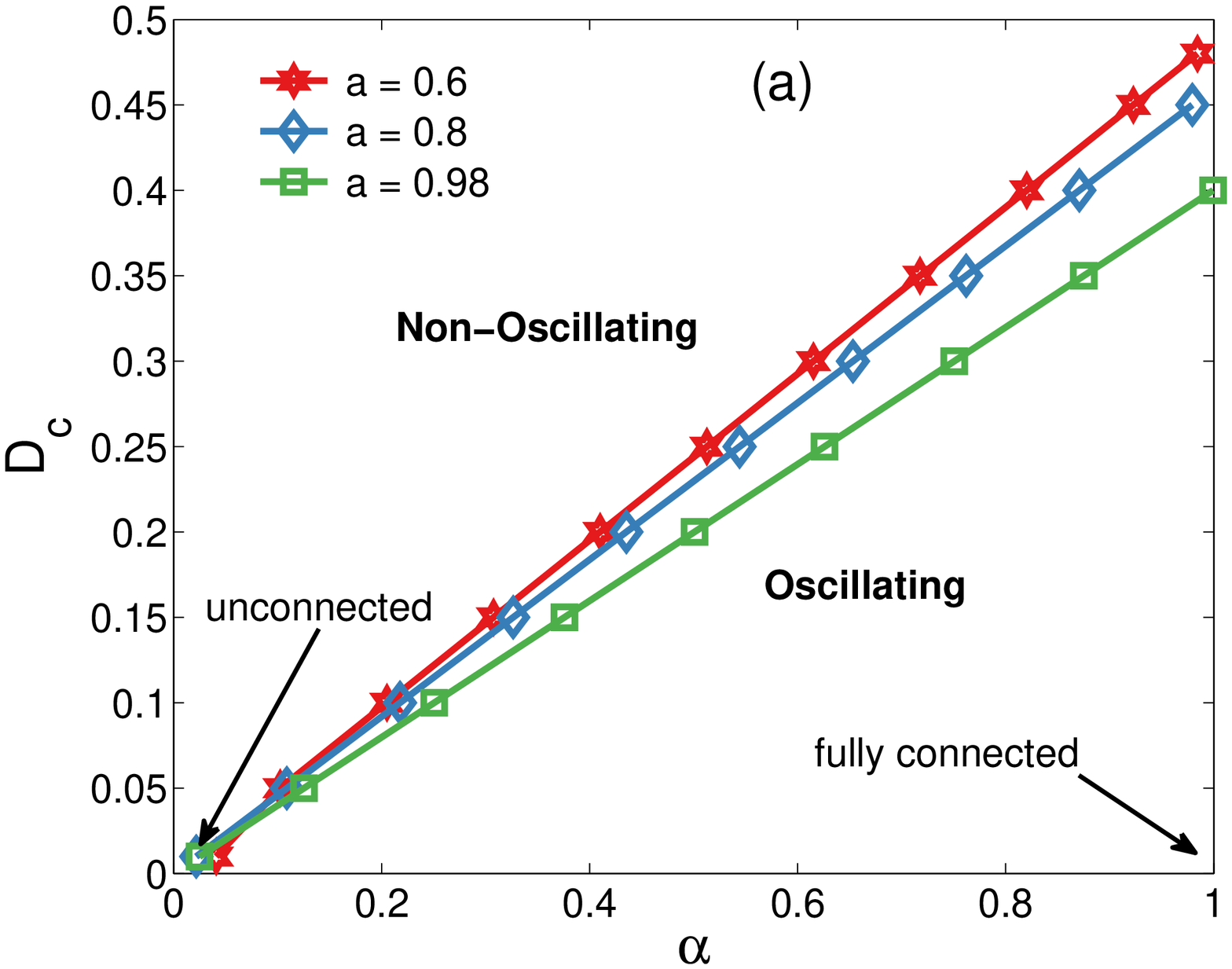}&
\includegraphics[width=0.49\linewidth]{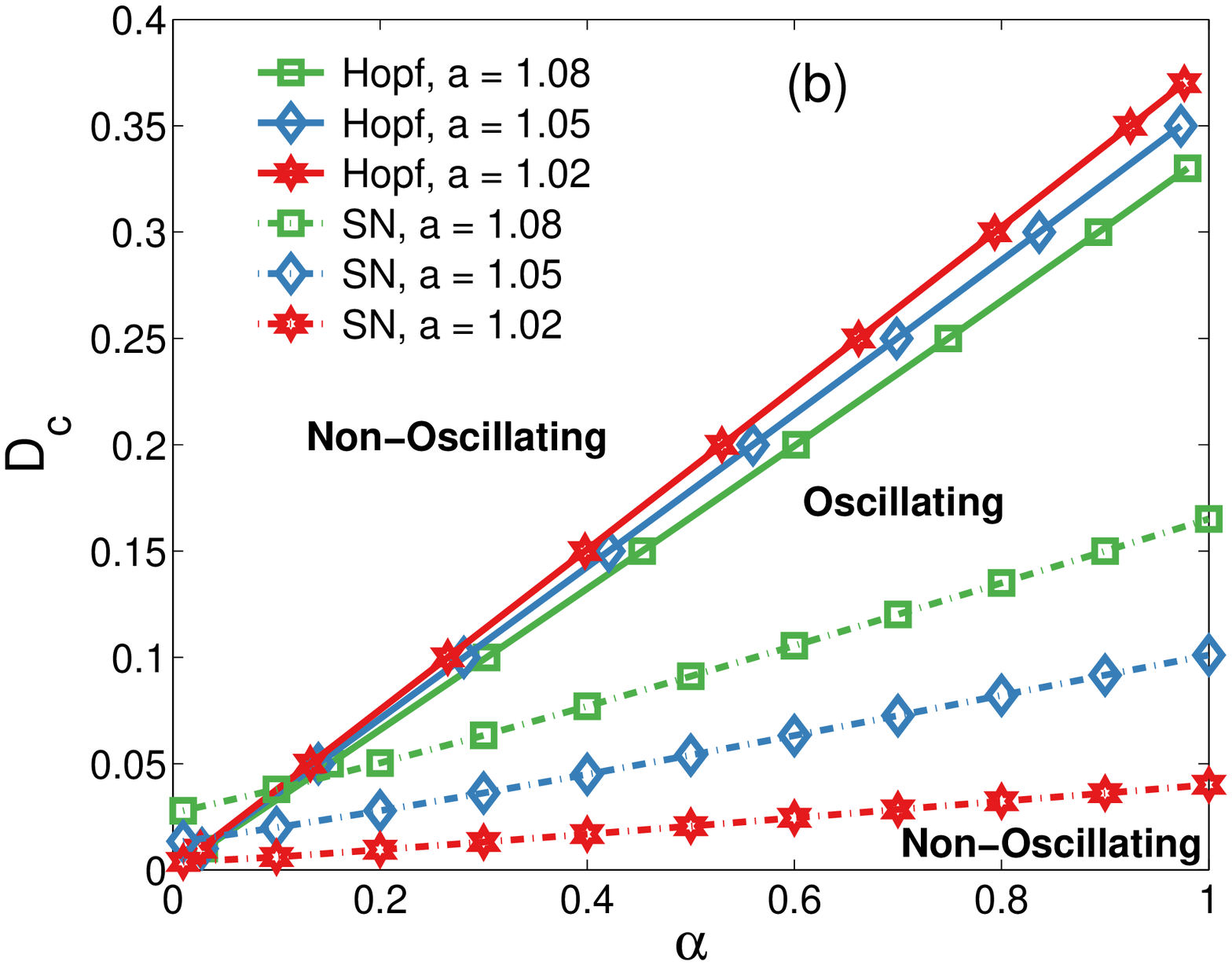}\\
\end{tabular}
\caption[]{(Color online) 
Critical noise intensity $D_c$ as function of the connectivity fraction $\alpha$.
(a): For low and moderate threshold $a$, only the Hopf bifurcations are present. (b): 
Large values of $a$ with Hopf and saddle-node (SN) bifurcations present.  Parameter values are indicated in the legends.
The coupling strength is $\kappa=1$ for both panels.}
\label{meank}
\end{figure}

We detect numerically the dependence on $\alpha$ of Hopf and saddle-node bifurcations 
in Eqs. \eqref{dotmvar_reg} for different excitation thresholds $a$
with the help of the software package MATCONT \cite{DhGoKuz03}. 
Results are summarized in Fig. \ref{meank} showing the bifurcation value of noise intensity (called the critical noise intensity) $D_c$ versus $\alpha$ for fixed values of the excitability parameter $a$.
We observe that both the Hopf and the saddle-node bifurcations are shifted to smaller noise intensities 
when the connectivity is decreased. 
Dependence of the Hopf threshold on $\alpha$ is practically linear,
in accordance with the expansion
\begin{equation}
D_c=\kappa\alpha\left(\frac{1}{2}-\frac{3}{32}a^4-\frac{3}{256}a^8+\ldots\right).
 \label{Dexpansion}
\end{equation}
As one would expect, the oscillatory regions shrink for higher values of threshold $a$.

\subsection{Binary random networks}
A simple example of a complex heterogeneous network is a binary random network which possesses two distinct connectivity degrees.
Denoting these as $k_1$ and $k_2$ ($\alpha_1=k_1/N$, $\alpha_2=k_2/N$), the degree distribution is given by
\begin{equation}
 P(k)=p \delta_{k,k_1}+(1-p)\delta_{k,k_2}.
\label{degreedist}
\end{equation}
We observe that these networks tend to be disassortative in the sense that nodes with different degrees are 
preferentially connected \cite{New02}. Such degree-degree correlations could probably be factored in by developing a theory along 
the lines of \cite{BoPast02}. It turns out (not shown here) that the disassortativity attains large values, if the variance among the connections is large. 
The maximum variance that can be obtained with \eqref{degreedist} is $\text{VarMax}=0.25$. We consider only values up to $\text{Var}(\alpha)/\text{VarMax}=0.5$, 
where the disassortativity is negligible.
\begin{figure}
\centering
\includegraphics[width=0.9\linewidth]{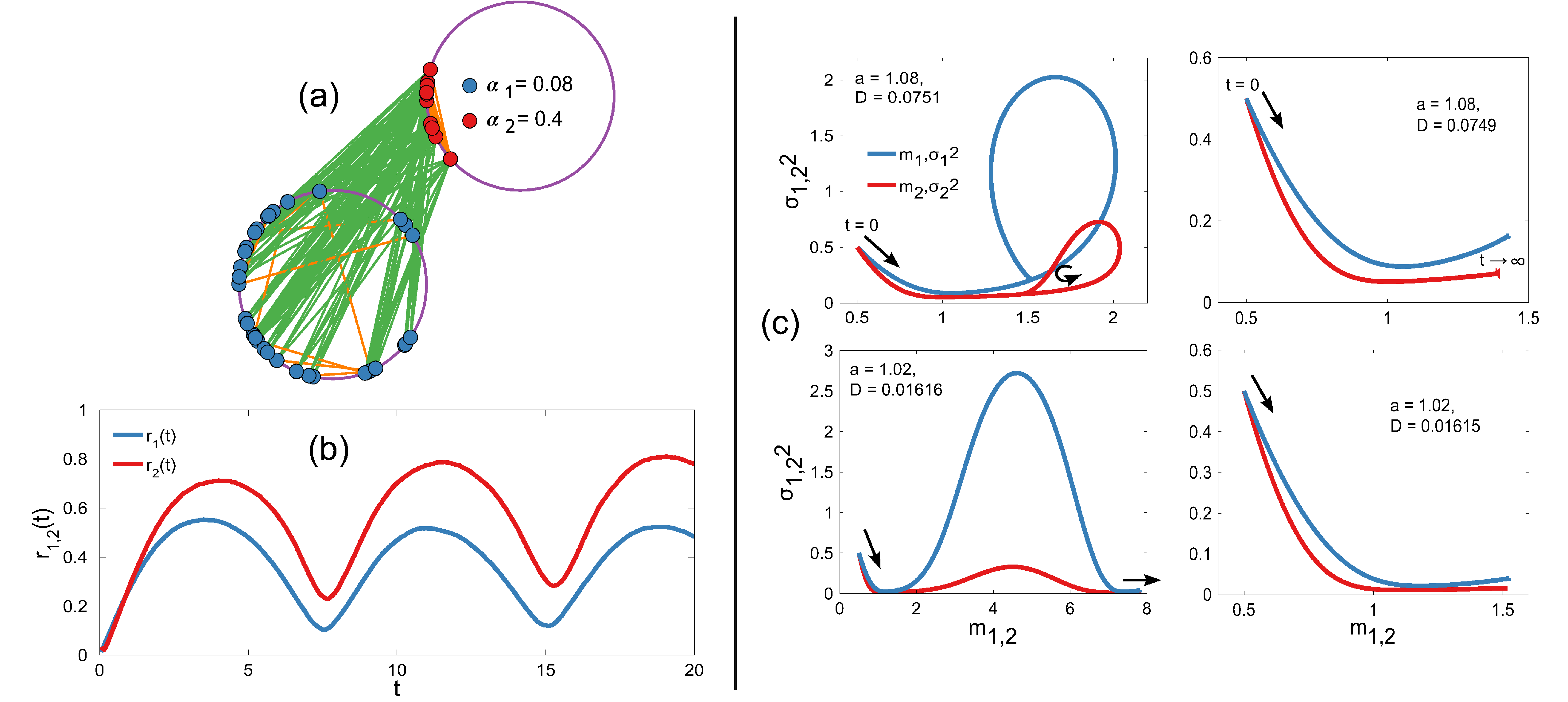}
\caption[]{(Color online) Dynamics of a binary random network of stochastic rotators. (a): A snapshot at $t=100$ of a network composed 
of $50$ elements. For clarity, the nodes are distributed on two separated rings according to their degrees and phases. Blue 
nodes have degree $k_1=4$, red ones have $k_2=20$. Green links are chosen between different degrees, while the orange lines connect 
nodes with the same degree. Remaining parameters are $a=0.2,\ D=0.1,\ \kappa=2,\ p=0.7$. (b): Order parameters 
$r_1(t)=\left|\rho_1(t|k_1)\right|$ and $r_2(t)=\left|\rho_1(t|k_2)\right|$ obtained from simulation of the {\em full dynamics}, 
Eq. \eqref{ourmodel}. The parameters are: $N=10^4,\ k_1=1000,\ k_2=4000,\ p=0.7,\ a=0.6,\ D=0.1,\ \kappa=1.5$. (c): Phase portraits 
of the GA Eqs. \eqref{mvar} for $\alpha_1=0.37$, $\alpha_2=0.97$, $p=0.95$, $\kappa=1$ and initial conditions 
$m_{1,2}(t=0)=\sigma_{1,2}^2(t=0)=0.5$. Upper left panel depicts localized oscillations, lower left panel depicts full-circle rotations.
In the right panels a small decrease of noise 
has the result that the system approaches fixed points (right endpoints of the trajectories).}
\label{rings}
\end{figure}
Inserting the degree distribution \eqref{degreedist} into Eqs. \eqref{dotmvar} gives a system of four coupled first-order differential equations:
\begin{equation}
 \begin{aligned}
\dot m_1=&1-\mathrm e^{-\sigma_1^2/2}\cosh(\sigma_1^2)\left[a\sin(m_1)+\frac{\kappa\alpha_1\alpha_2 (1-p)}{p\alpha_1+(1-p)\alpha_2}\sin(m_1-m_2)\mathrm e^{-\sigma_2^2/2}\right] \\
\dot \sigma_1^2=&2D-2\mathrm e^{-\sigma_1^2/2}\sinh(\sigma_1^2)\\
&\times\left[a\cos(m_1)+\frac{\kappa\alpha_1}{p\alpha_1+(1-p)\alpha_2}\left(\alpha_1 p\mathrm e^{-\sigma_1^2/2}+\alpha_2 (1-p)\mathrm e^{-\sigma_2^2/2}\cos(m_1-m_2)\right)\right] \\
\dot m_2=&1-\mathrm e^{-\sigma_2^2/2}\cosh(\sigma_2^2)\left[a\sin(m_2)+\frac{\kappa\alpha_1\alpha_2 p}{p\alpha_1+(1-p)\alpha_2}\sin(m_2-m_1)\mathrm e^{-\sigma_1^2/2}\right] \\
\dot \sigma_2^2=&2D-2\mathrm e^{-\sigma_2^2/2}\sinh(\sigma_2^2)\\
&\times\left[a\cos(m_2)+\frac{\kappa\alpha_2}{p\alpha_1+(1-p)\alpha_2}\left(\alpha_2 (1-p)\mathrm e^{-\sigma_2^2/2}+\alpha_1 p\mathrm e^{-\sigma_1^2/2}\cos(m_2-m_1)\right)\right].
 \end{aligned}
\label{mvar}
\end{equation}
Bifurcation analysis of this system was performed using MATCONT \cite{DhGoKuz03}.

Fig.~\ref{rings}(a) illustrates a binary random network. In (b) we show the order parameters $r_{1,2}(t)$ \eqref{rkthetak} obtained 
by simulation of the full dynamics \eqref{ourmodel}. 
More connections appear to increase the similarity between the 
phases in the corresponding population. Fig. \ref{rings}(c) shows phase portraits 
of the GA Eqs. \eqref{mvar} near the saddle-node bifurcation. The left panels 
depict the oscillatory regime. Indeed, the set of nodes with more connections is more synchronized both in the case of localized oscillations
(upper left) and for full rotations (lower left). This is signalled by smaller variances $\sigma_2^2(t)$. 
In the right panels of (c) a tiny decrease in the noise intensity brings the whole system into a stationary state, where 
the variables $m_{1,2},\ \sigma_{1,2}^2$ approach a fixed point. 

In this context we never observed that one population of a certain degree 
is oscillating, while the other is not. This might be the result of entanglement of two populations with different connectivity degrees. Hence,
a visualization such as in Fig.~\ref{rings}(a) has to be taken with care.

\begin{figure}
\centering
\begin{tabular}{cc}
\includegraphics[width=0.49\linewidth]{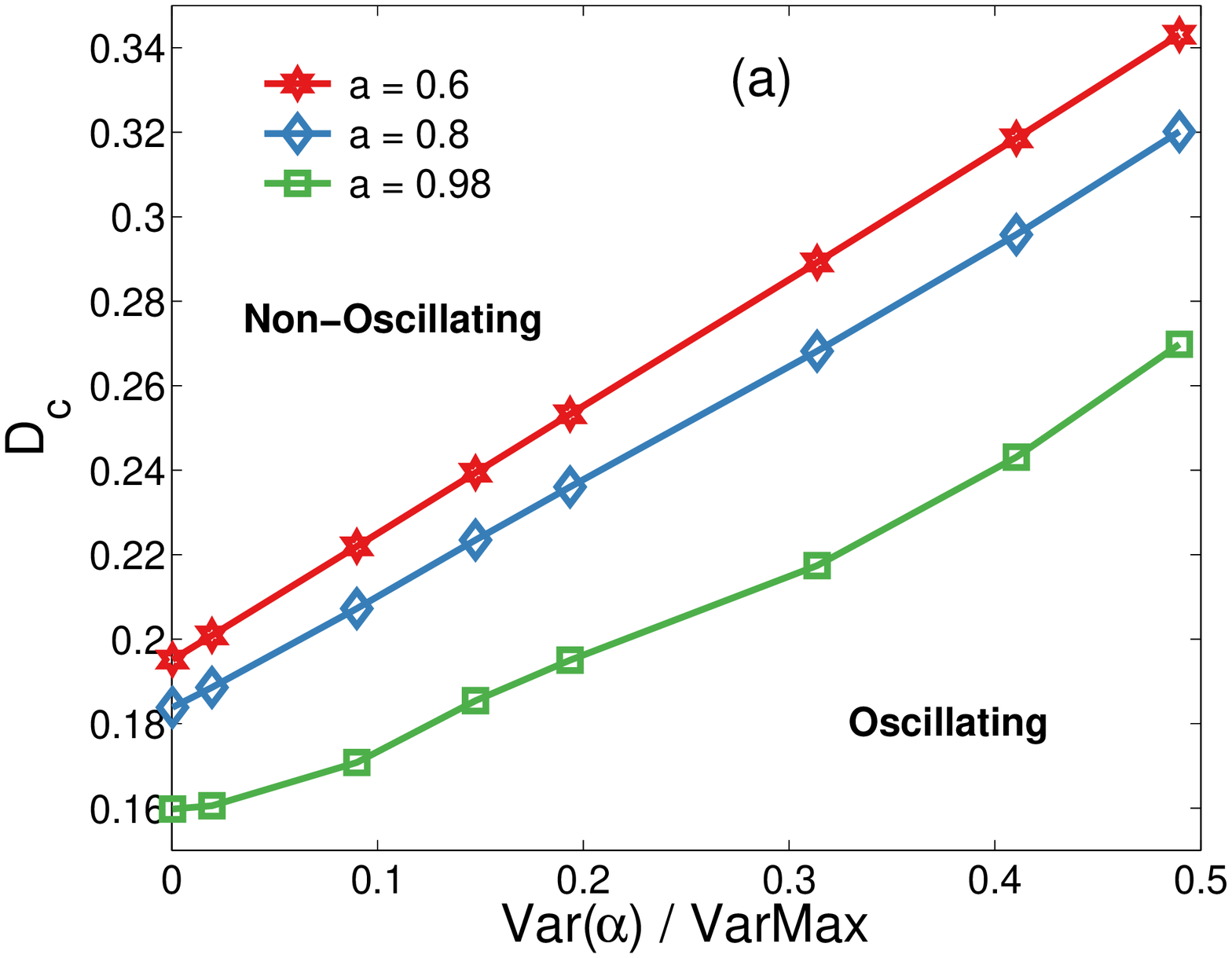}&
\includegraphics[width=0.49\linewidth]{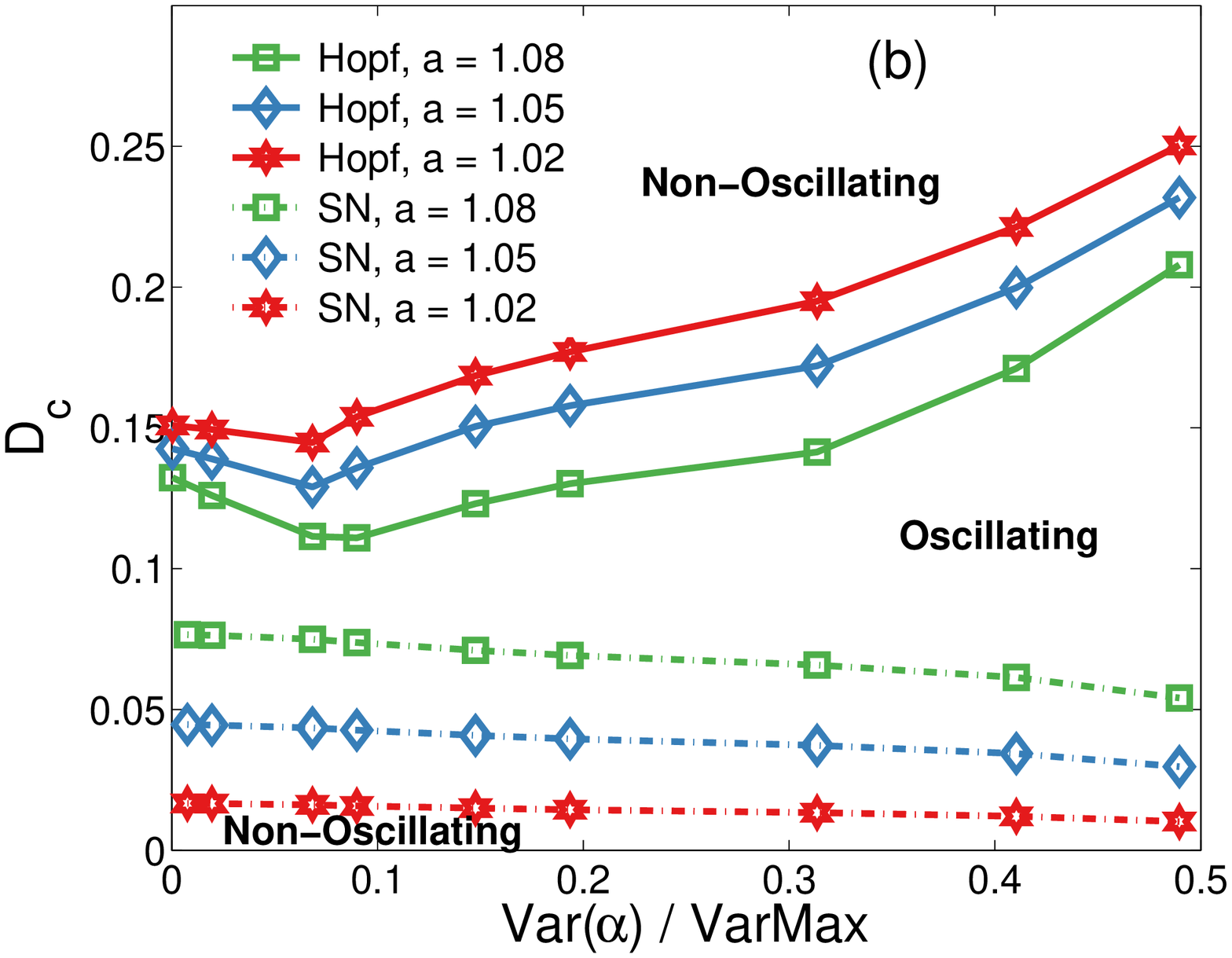}\\
\end{tabular}
\caption[]{(Color online) Critical noise intensity $D_c$ versus $\text{Var}(\alpha)/\text{VarMax}$ obtained in the GA for the binary random network of stochastic  rotators.  As in Fig. \ref{meank} panels (a) and (b) show the oscillatory and the excitatory regimes, respectively. 
The values of excitability parameter $a$ are indicated in the legends along with the type of bifurcation: Hopf or saddle-node (SN). Other parameters are
$\text{VarMax}=0.25$, $\langle \alpha \rangle=0.4$, $\kappa=1$.}
\label{vark}
\end{figure}

In order to isolate the effects of network heterogeneity, we fix the average degree,
$\langle\alpha\rangle=0.4$. The variance of $\alpha$ is tuned by varying three parameters: $\alpha_1$, $\alpha_2$ and $p$. Since $\text{Var}(\alpha)$ is not uniquely defined and different parameter sets yield the same variance, we take the path where the $\alpha$-values are the largest, because then the mean-field assumption is best fulfilled. 
Fig.~\ref{vark} shows critical noise intensity $D_c$ at which a Hopf or saddle-node bifurcations occur in the GA system (\ref{mvar}) versus the connectivity variance $\text{Var}(\alpha)/\text{VarMax}$.
In the oscillatory regime, $a<1$, shown in Fig.~\ref{vark}(a),  $D_c$ vs $\text{Var}(\alpha)/\text{VarMax}$ shows trends similar to the case of $a=0$ \cite{SonnSchi12},
\begin{equation}
D_c(a=0)=\frac{\kappa}{2N}\frac{\langle k'^2 \rangle}{\langle k'\rangle}=\frac{\kappa}{2}\left(\langle\alpha\rangle+\frac{\text{Var}(\alpha)}{\langle\alpha\rangle}\right).
\end{equation}
Nonlinear dependence of $D_c$ vs $\text{Var}(\alpha)$ starts to appear only near the threshold $a=1$. 
In the excitable regime $a>1$ [Fig.~\ref{vark}(b)] the oscillatory region grows with the increase of $\text{Var}(\alpha)$, if the variance is already large. In particular, the Hopf bifurcations shift to larger noise values, while the saddle-node bifurcations shift to smaller ones. 
Furthermore, for the Hopf bifurcation the dependence $D_c$ vs $\text{Var}(\alpha)$ becomes non-monotone with minima at 
moderate values of the structural heterogeneity. These minima become more pronounced for larger values of $a$. This effect constitutes a qualitative difference to the oscillatory regime, $a<1$, 
and is induced by the network structure.

In particular, the minima observed for the Hopf bifurcation lines result from adding a small fraction of highly connected nodes, so-called hubs. If the nodes with larger degree are in the majority, then the critical noise is larger
and the minima disappear (not shown here).

\subsection{Numerical simulations of binary random networks} 
To test theoretical predictions made with the mean-field theory in Gaussian approximation, we performed numerical simulations of random binary networks of stochastic rotators. 
The goal was to check whether the global oscillations can indeed be suppressed by increasing the network heterogeneity from a small to a moderate value. We took networks 
of $N=10^4$ rotators and integrated Eq. \eqref{ourmodel} using the Heun method \cite{Man00} with the time step of $0.5$. Smaller time steps down to $0.05$ had no significant effect on 
the simulation results.

We used the Viger-Latapy algorithm to generate the binary random networks \cite{ViLa05}. In particular, we used the $\texttt{C}$ library $\texttt{igraph}$ \cite{CsNe06} (version $\texttt{0.6.5}$),
where the aforementioned algorithm is implemented.

\begin{figure}
\centering
\includegraphics[width=0.99\linewidth]{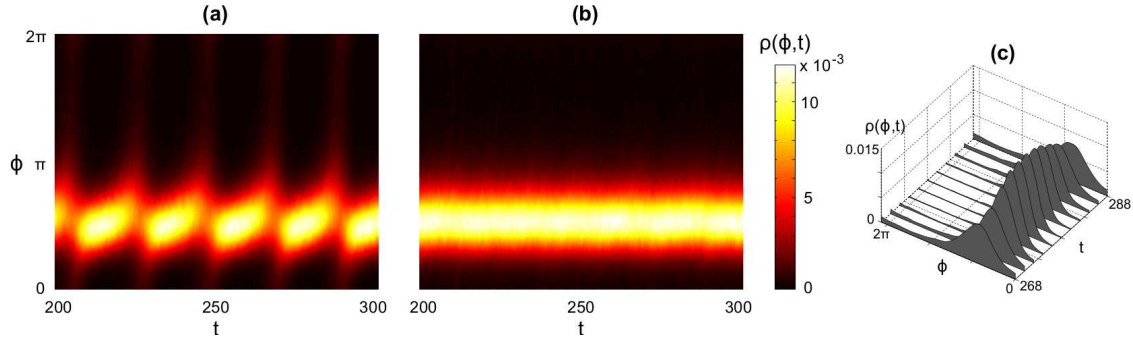}
\caption[]{(Color online) Numerical simulations of random binary networks of $N=10^4$ stochastic rotators. Shown are time-dependent probability densities of rotators phases, $\rho(\phi,t)$. 
In both panels the rotators are in the excitable regime with $a=1.05$, $D=0.095$ and $\kappa=2$.
(a): Homogeneous binary network with parameters $\alpha_1=0.59$, $\alpha_2=0.39$ and $p=0.05$ leading to the variance $\text{Var}(\alpha)/\text{VarMax}=0.0076$.
(b): Inhomogeneous binary network, $\alpha_1=0.97$, $\alpha_2=0.37$, $p=0.05$ with the variance $\text{Var}(\alpha)/\text{VarMax}=0.0684$.
(c): Projections from the phase distribution in (a) for one period.}
\label{rho_var}
\end{figure}

Fig.~\ref{rho_var} shows simulation results for two different networks. Both networks had the same average degree, but distinct variances.
Compared with Fig.~\ref{vark}, the only difference in parameters of Fig.~\ref{rho_var} is a larger value of the coupling strength which has no qualitative effect on global dynamics but simplifies the detection of the global oscillations. 
As can be seen from panels (a) and (b), an increase of the network heterogeneity leads to suppression of network's oscillations, in agreement with the theory. 
Panel (c) shows snippets of the phase distribution with the parameters from panel (a). As in Fig. \ref{rings}(c)[upper left], we
observe localized oscillations. One can recognize the distinct single-hump shape that is necessary for the Gaussian approximation (section \ref{gaussian_approx}).

\section{Conclusion}
\label{conclusion}
We have studied a well-known model for excitable dynamics, the so-called 
active rotators, in the context of heterogeneous networks. 
Besides investigating the effects of noise, 
we put emphasis on the effects that emerge due to 
different coupling structures. Considering the infinite system-size limit,
the first step in our analysis has been to reduce the dimensionality of the system. 
Coarse-graining of the network has lead us to the mean-field description. 
Assuming further that the phases in each set of oscillators with the same degree 
obey a Gaussian distribution with time-dependent cumulants, we arrived at 
a lower-dimensional set of integro-differential equations.
The order of this system equals the doubled number of different degrees
occuring in the network.
The approach has been used to perform numerical bifurcation
analysis on regular and binary random networks. 
The calculated state diagrams display the parameter regions 
where the neuronal network shows synchronous or asynchronous spiking behavior. 
Transitions between these regions are typically related to passages through
the Hopf and saddle-node bifurcations. 
When the noise intensity is increased from zero, the saddle-node bifurcations 
mark the onset of global firing in the ensemble, whereas the Hopf bifurcation 
leads to the disappearance of the collective mode of synchronous 
oscillations. 
We have been able to show that changing the network structure alone,
without variations in the noise intensity and/or the excitation threshold,
is sufficient for a switch between the different patterns of spiking behaviors. 
In fact, adaptations in the coupling structure play a crucial role 
for the information processing in populations of neurons \cite{Sp10}.

Our results suggest that deleting uniformly connections of all nodes in the network
does not lead to new characteristics, 
since the coupling strength is only rescaled effectively. 
However, if nodes differ in their number of connections, 
an interesting phenomenon is obtained in the excitable regime. 
We have found out that the critical noise intensity,
at which the Hopf bifurcation occurs and small-scale oscillations are born, 
can possess a minimum for a moderate network heterogeneity. 
It was induced by the presence of a small fraction of highly connected nodes. 
This might have meaningful implications for the functioning of neuronal networks,
since such hub nodes naturally exist \cite{Sp10,VarChPanHaChk11}.
It is well-known that neuronal networks are highly 
heterogeneous. Our analysis further suggests that this property 
implies existence of large parameter regions to get global firing. 
If, however, a specific task requests a switching between different spiking patterns, reducing the heterogeneity in the connectivities becomes beneficial, since 
only small noise variations are needed. The optimum for such a task would then be given by a moderate coupling variance with a small fraction of hub nodes.

\begin{acknowledgement}
This work was supported by the GRK1589/1 and project A3 of the Bernstein Center for Computational Neuroscience Berlin. Research of M.Z. 
was supported by the project D21 of the DFG Research Center MATHEON.
\end{acknowledgement}

\bibliography{bibliography}
\bibliographystyle{unsrt}
\end{document}